# Python for Smarter Cities: Comparison of Python libraries for static and interactive visualisations of large vector data

Gregor Herda[a] *, Robert McNabb[a]

[a] Ulster University, School of Geography and Environmental Sciences, Coleraine, UK

* Corresponding author: Gregor Herda; herda-g@ulster.ac.uk

**Abstract**

Local governments, as part of 'smart city' initiatives and to promote interoperability, are increasingly incorporating open-source software into their data management, analysis, and visualisation workflows. Python, with its concise and natural syntax, presents a low barrier to entry for municipal staff without computer science backgrounds. However, with regard to geospatial visualisations in particular, the range of available Python libraries has diversified to such an extent that identifying candidate libraries for specific use cases is a challenging undertaking. This study therefore assesses prominent, actively-developed visualisation libraries in the Python ecosystem with respect to their suitability for producing visualisations of large vector datasets. A simple visualisation task common in urban development is used to produce near-identical thematic maps across static and an interactive 'tracks' of comparison. All short-listed libraries were able to generate the sample map products for both a small and larger dataset. Code complexity differed more strongly for interactive visualisations. Formal and informal documentation channels are highlighted to outline available resources for flattening learning curves. CPU runtimes for the Python-based portion of the process chain differed starkly for both tracks, pointing to avenues for further research. These results demonstrate that the Python ecosystem offers local governments powerful tools, free of vendor lock-in and licensing fees, to produce performant and consistently formatted visualisations for both internal and public distribution.

**Keywords:** Python, open-source, visualisation, vector data, Smart City

## 1   Introduction

The 'smart city' paradigm hinges on the idea that modern information and communications technology can be harnessed to 'automate routine functions serving individual persons, buildings, [and] traffic systems' and 'enable us to monitor, understand, analyse and plan the city to improve the efficiency, equity and quality of life for its citizens in real time' (Batty et al., 2012). While the term 'smart city' remains a fuzzy concept, the literature broadly agrees that approaches under this umbrella can 'contribute to the legitimacy of urban governance' by helping improve outcomes of decision-making and strengthen democratic forms of government through knowledge creation, knowledge exchange and innovation (Meijer and Rodrigues Bolivar, 2015).

A crucial condition for fulfilling this promise will be the ability of local governments to analyse ever larger volumes of data and move from reactive to predictive, targeted service delivery (Al Nuaimi et al., 2015; Davenport et al., 2012; Malomo and Sena, 2016). In the UK, measures related to expanded use of dedicated and shared data analysis capacities leading to better allocation of resources was estimated to yield public sector savings of between GBP10 billion and GBP23.4 billion over five years (Copeland, 2015). Considering the post-2020 "scissors effect" of increasing expenditures and declining revenues (OECD, 2020), stakeholders will need to find new ways to realise these savings.

In recent years, examples of innovative applications of data science by, or simply "in", cities have proliferated, ranging from more accurately estimating neighbourhood incomes (Glaeser et al., 2015); improving solid waste management (Watson and Ryan, 2020); or providing night-time public transport (Hong et al., 2019). A review by De Souza et al. (2019) found predictive analytics to be the most fertile application for machine learning in the 'smart city' context.

Several studies have pointed out that scaling up these initiatives will not be easy, especially in a likely "do-more-with-less" resource environment. While some organisations have yet to optimally manage and de-silo the 'small data' they currently possess (Copeland, 2015; Malomo and Sena, 2016; Oates, 2018; Oxford Internet Institute, 2019; Ross et al., 2013), others will need to find ways to exploit cross-organisational economies of scale by reducing duplication of hardware, systems and personnel (Copeland, 2015), and all will have to address questions of inequitable data access, biased data collection, and politicisation of 'smartness' itself (Kitchin, 2014; Watson, 2013; Zanmiller, 2015). Finally, developing the technical and human capacities for more advanced urban analytics will only be an intermediate step towards efficiently structuring these capacities into teams and departments (Günther et al., 2017) and reforming governance structures to respond to data-driven decision-making (Sharma et al., 2014).

In any case, what should precede reforms in any data domain relevant to making cities 'smart*er*', from data acquisition and management, to analysis and visualisation, is an objective evaluation of available tools and their respective barriers to entry. In this study, open-source data visualisation as an important step in the data value chain is used to illustrate such an evaluation.

The paper first presents the context in which local governments may wish to explore open-source approaches to data visualisation as well as the challenges this may pose. It then presents an overview of the Python visualisation landscape with a focus on geospatial applications. A common visualisation task was developed and implemented using Python libraries across two separate competitive 'tracks', producing static and interactive visualisations, respectively. Both 'soft' performance indicators, such as available project documentation and code complexity are examined alongside quantitative comparative indicators such as output file size and cumulative CPU runtime. Finally, implications for local authorities arising from engaging with the underlying open-source business model are outlined.

## 2   New approaches to data visualisation for local government

As limitations on bandwidth and server capacity are lifted, geospatial visualisations are, in many contexts, moving beyond the provision of image tiles (Stefanakis, 2017) towards high-resolution, on-demand, and even real-time representations. With local governments both gathering, and becoming aware of the utility of, ever-larger amounts of data, the need increases for 'new approaches to an efficient, consistent, and compliant mapping of massive polyline geometries representing geographic features or trajectories' (Breunig et al., 2020).

Given increasing time demands on municipal workers (Oxford Internet Institute, 2019), it is not implausible to assume that the more intuitive these new approaches are to master and apply, the more likely they are to take hold in local government.

Obviously, staff training will only be part of the solution. Another will be the identification and development of appropriate systems satisfying an expansive list of requirements (Daneva and Lazarov, 2018). The question is, which systems for which services? Concerning software solutions, some municipal functions can and certainly will be supported by proprietary solutions, though a growing number of local governments are choosing open-source alternatives (European Commission, 2021). In the realm of data management, analysis and visualisation, the capabilities of, among others, PostgreSQL, R, and Python offer an increasingly competitive value free of vendor lock-in and licensing fees (Burger, 2019; Oxford Internet Institute, 2019).

Python in particular presents low barriers to entry to newcomers due to its concise, natural syntax. It is also a highly innovative ecosystem. Suites of related functionalities are bundled into hundreds of 'packages' or 'libraries' distributed to users on request or automatically as part of other libraries' dependencies. Due to Python's open-source nature, each of these libraries can leverage functionalities from another in the same production environment without data having to be mediated through files (Lin, 2012). This enables organisations dealing with an ever-expanding kaleidoscope of data sources and consumers to establish lean data manipulation and management workflows.

A first barrier to uptake is the challenge of identifying suitable libraries. The questions which this study seeks to answer on behalf of new organisational users are:

1. Does this library support the type of map product we wish to produce?
2. What kind of input and output data formats are supported, and is elaborate pre-processing required?
3. Can large datasets be visualised while retaining high performance, both during development and production?
4. How complex are its underlying concepts and syntax, and what resources are available to master it?
5. Is the developer and user community active and stable, ensuring that the library will be supported well into the future?

## 3   The geospatial Python visualisation landscape

In recent years, the number of Python libraries with geospatial functionalities has multiplied significantly. **Error! Reference source not found.**, an expansion on VanderPlas (2017), presents a non-exhaustive overview of some of the most prominent Python libraries involved in data visualisation. The landscape is centred around three main clusters: primarily static visualisations building on *Matplotlib* (Barrett et al., 2005; Hunter, 2007), interactive visualisations interfacing with HTML and *JavaScript*, and advanced 2D/3D visualisations and animations relying on the *OpenGL* application programming interface (API). Libraries included in this study's long list are highlighted.[1]

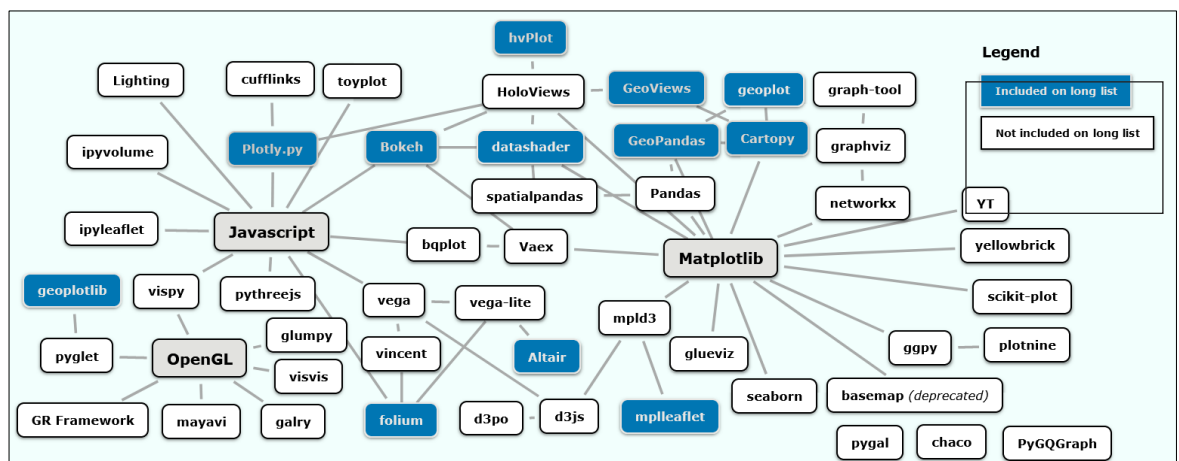

**Figure 1:** The Python visualisation landscape, adapted from VanderPlas (2017)

*Matplotlib* has been the de-facto standard for scientific plotting for nearly two decades. Numerous other libraries build on *Matplotlib* and try to expand on and/or simplify its complex, highly customisable, functionalities (see examples in Matplotlib, 2021). *Matplotlib*, in combination with other libraries, is also capable of rendering geospatial data. Initial milestones in this achievement were the translation of common GIS formats into Python mappings (Gillies, 2011) which enabled the spatialisation of the popular *Pandas* data analysis library (Stancin and Jovic, 2019) through the implementation of *Shapely* geometries by *GeoPandas*, a library with a central position for geospatial operations in Python (Jordahl, 2014). Since its inception, *GeoPandas* has provided a high-level interface with *Matplotlib* (GeoPandas, 2021). The early to mid-2010s also saw an ever-increasing range of libraries interface with the JavaScript or OpenGL clusters to produce interactive graphs.

Across all clusters, libraries differ between *imperative* or *declarative* visualisation approaches: in an imperative approach, the user manually specifies *how* something should be rendered, including individual plotting steps; specification and execution are intertwined. In a declarative approach, the user only specifies *what* should be rendered, with plotting details determined automatically. This separates specification from execution (VanderPlas, 2017).

## 4    Data and methods

*4.1    General approach, data source and software environment*

This study sought to identify visualisation libraries in the Python ecosystem suitable for visualising 'large' datasets, while at the same time providing a relatively easy to learn syntax, a vibrant developer and user community, and reasonable performance.

As static and interactive visualisations differ in the way they are produced, distributed, and consumed (Bednar 2018a), the comparison was conducted along two separate tracks: static and interactive, with libraries in each track representing both the imperative and declarative approach.

In both tracks, a map 'template' following a uniform symbology with standardised map elements was reproduced, displaying two tables from the complete city of Dresden's real-estate cadastre, and a smaller subset (Table 1).

**Table 1:** Spatial tables selected from Dresden's real estate cadastre

| Original table name | Feature type | Geometry type | Feature count (complete) | Feature count (subset) |
|---|---|---|---|---|
| *AX_Gebaeude* | Buildings | CurvePolygon | 143,575 | 2,645 |
| *AX_Fliessgewaesser* | Rivers or streams | CurvePolygon | 1,152 | 12 |

Each dataset was converted to a separate locally-hosted PostGIS v3.0.0-enabled PostgreSQL v12.4 database. Through the Visual Studio Code (VSCode) v1.59.1 Python Extension, data was loaded into a Python v3.8.8. kernel via *GeoPandas'* `from_postgis()` function. Further details on system specifications, the data source and pre-processing steps are outlined in the *Data and methods* section of the Supplementary Material.

---

[1] Libraries shown here can be broadly grouped into two categories: core plotting libraries performing the bulk of the visualisation directly, often by relying on more fundamental low-level technologies (e.g. *Matplotlib, Bokeh, Vega* or *pyglet*), and libraries offering high-level interfaces to the former (e.g. *Pandas* and its derivatives, HoloViz products such as *hvPlot* or *GeoViews,* or *Altair* and *geoplotlib*).

*4.2 Library selection*

To gain an overview of candidate libraries, a "long list" based on the list collated by PyViz.org (2019) was compiled, including metadata such as:

1. Output formats;
2. general implementation strategy;
3. Installation channels and requirements;
4. Input formats and required data conversions;
5. Proxies suggesting the vibrancy of the developer and user community.

The complete long list for both tracks is presented in section S1.2 and Tables S1 and S2 of the Supplementary Material.

The long list confirms that limited support for input formats is not a concern. Through *GeoPandas*, users gain access to GDAL/OGR's 90 vector and 166 raster drivers (GDAL, 2021a, 2021b), with most libraries able to process GeoDataFrames either directly or after at most two conversion steps.

Based on this long list, for each track a final short list (Table 2) consisting of five libraries or library combinations (hereafter referred to as 'implementations') was created.

| Static | | Interactive | |
|---|---|---|---|
| **Imperative** | **Declarative** | **Imperative** | **Declarative** |
| *GeoPandas* v0.9.0 (Jordahl et al., 2021) + *Matplotlib* | *datashader* v0.13.0 (Bednar et al., 2021) | *Bokeh* v2.3.2 (Bokeh Development Team, 2018) | *GeoViews* + *Bokeh* |
| *Cartopy* v0.18.0 (Met Office, 2021) + *Matplotlib* | *Altair* v4.1.0 (VanderPlas et al., 2018) + *Vega-Lite* (Satyanarayan et al., 2017) | *GeoViews* v1.9.1 (Rudiger et al., 2021a) + *datashader* + *Bokeh* (Bokeh Server) | *hvPlot* v0.7.3 (Rudiger et al., 2021b) + *GeoViews* + *Bokeh* |
| *geoplot* v0.4.3 (Bilogur et al., 2021) + *Matplotlib* | | *Plotly.py* v4.14.3 (Plotly Technologies Inc., 2015) | |

**Table 2:** Library short list

The short list tried to include both large-community projects (e.g., *Bokeh* and *Plotly*) as well as libraries relying on a more limited number of contributors (e.g., *geoplot*). All short-listed libraries needed to show on-going development in the last year (hence the exclusion of *mplleaflet* and *geoplotlib*) and needed to be able to plot geometries without employing a Web Tile Service (hence the exclusion of *folium*). A variety of backends and both imperative and declarative approaches are included. Finally, *datashader* was included in the static track 'out of competition', to first demonstrate its 'as-is' functionality before employing it in the interactive track in conjunction with a core plotting library.

*4.3 Library comparison*

The short-listed libraries were compared with regard to:

1. the range of available documentation, based on common documentation 'elements' and code examples consulted to implement the task;
2. the number of lines of code needed to reproduce the map template, excluding comments and blank lines. This provided a proxy measure for code complexity, using a 'reduced' version of the more feature-rich extended scripts. The reduced versions do not wrap rendering-related assignment statement in a function (the `renderFigure()` function of the extended scripts) and exclude optional user inputs. To improve code legibility and comparability, the same level of intermediate assignment statements across libraries were used.[2];

3. the ability to reproduce the map template including a legend and base map;
4. resource requirements (output file size after export to SVG/PNG or HTML; for interactive visualisations, a subjective assessment of 'responsiveness' on pan and zoom);
5. the cumulative CPU runtime required for the Python-based portion of the main `renderFigure()` function to complete, indicated as an average across a total of 10 runs. To increase comparability, the rendering function excluded both data acquisition and any data pre-processing, reprojection or conversion steps. CPU runtimes were measured using the *cProfile* module (Python Software Foundation, 2021) before writing the results of individual runs to labelled *cProfile* output files in the binary `.prof` format for analysis. As rendering is ultimately performed by the execution of JavaScript code, *cProfile* will not capture the entirety of the processing costs for the interactive track. See section S1.2 of the Supplementary Material for details on measures taken to enhance comparability of CPU runtimes. All scripts and `.prof` files are provided as part of a public GitHub repository (Herda, 2022).

*4.4 Study limitations*

Limitations for two of the chosen indicators should be highlighted here. Employing any of the short-listed libraries involves multiple complex technologies, some of which exist outside the Python ecosystem (e.g., Jupyter Notebook or the JavaScript libraries underlying both *Bokeh* and *Plotly.py*). Due to these confounding variables, the various adjustments in Table S3 are unlikely to have established a truly level playing field with regard to comparing CPU runtimes. What the *cProfile* results seek to demonstrate is a *relative* comparison of the user experience and the approximate time required to generate a map product on screen.

Due to the subjectivity of lines of code as a measure for code complexity, an alternative methodology could see groups of experts for each implementation develop a 'best practice' code sample, though this may further reduce comparability.

It must also be acknowledged that the static map products would not be considered useful outputs in most real-world scenarios: due to significant overplotting, a building-level analysis would rarely be presented at such a small representative fraction. Additionally, when converting figures to SVG on the static track, the large file sizes using the complete dataset would cause most viewing applications to become unresponsive. The same is true for saving to HTML on the interactive track, as noted by Cuttone et al. (2016). Except for live server-side aggregation and rasterisation as demonstrated by the *GeoViews + datashader + Bokeh Server* implementation, most workflows targeting web-based distribution would involve prior conversion to a series of map tiles to then be served by client-side Web Tile Services (see, e.g., Morgan and Lovelace, 2020).

Similarly, the simplicity of the chosen visualisation task also meant that more advanced, and for local authorities potentially more interesting, use cases such as dashboarding were not demonstrated. Both *Bokeh's* and *Plotly's* native dashboarding capabilities as well as HoloViz's *Panel* library are therefore mentioned here.

---

[2] It should be noted that the degree to which new Python objects are created by way of an assignment statement to then be referenced later, instead of making them a key or positional argument directly, remains a question of personal choice.

Finally, due to the interconnectedness of the Python ecosystem, a library's functionalities and performance cannot be solely attributed to its own codebase. All libraries as well as their dependencies are under constant development. As such, the state of play outlined here represents merely a snapshot.

## 5 Results and discussion

### 5.1 General notes on library documentation and learning pathways

The comprehensiveness of official documentation and community-provided examples is arguably the most significant determinant in how quickly new users pick up underlying concepts and apply a library's functionality to a particular use case. Over the years, the structure and content of Python documentation has coalesced into a near-standard set of 'elements'.

A "quick start" or "getting started" section can be found in most libraries. These are usually structured as explanatory text alternating with sample code . Libraries targeting a broad data science audience, rather than the geospatial community, typically visualise simple mathematical functions in cartesian space (Altair-Viz, 2021). New learners without a background in data science or mathematics may find this an initial hurdle, especially once more advanced features are demonstrated using more complex functions (e.g., the data sample provided by Anaconda, Inc., 2021). Finding the right range and difficulty of examples for students of different backgrounds is a known problem in data science education (Hardin et al., 2015; Pournaras, 2017), and Python-based data visualisation is no exception.

While "getting started" sections are available as static web pages, they are often conversions of Jupyter Notebooks which allow users to edit and execute code themselves in a locally-run Python kernel. This popular channel for teaching Python, though superficially 'interactive', follows the vein of *behaviouristic* learning theory where teaching content is divided into small learning units and learners' initial role is to 'passively receive knowledge and follow instructions without involving peers in the learning process' (Zendler, 2019).

A second element present in all short-listed libraries were more extensive 'user guides'. These often covered advanced features and concepts of increasing complexity.

Thirdly, all libraries, except *hvPlot*, featured a so-called "API reference". This can be thought of as the official 'code registry' or 'code index', listing all functions and classes accessible to users, including requirements for positional or keyword arguments. API references largely reproduce the "docstrings", or non-code explanatory text, appended to a function, method, or class definition in Python code. Except for more advanced users, the API reference rarely serves as the first learning resource users engage with, but it is essential for exploring more advanced features.

Users will also find "example galleries" or "reference galleries" a helpful resource. These provide output examples along with sample data and underlying code. In an ideal case, they can significantly flatten the learning curve by requiring only minor adjustments to adapt to one's particular use case.

The formal documentation will not always be sufficient to help users understand underlying concepts or implement a particular use case. In these situations, two powerful community channels exist to help learners. The first is a library's GitHub repository, which allows for creating either an 'issue' (e.g., feature request or bug report) or a 'discussion' (e.g., for usage questions or general feedback). The second channel is appropriately tagged questions on the website StackOverflow (www.stackoverflow.com), which are often answered by members of a library's core developer team. These community-based resources expand the learning pathway towards the *cognitivist* approach, with learners actively organizing and reorganizing information, and peers being involved in, and influencing, the learning process (Zendler, 2019). Here, the Python community's "peer dynamic" exhibits similarities to beneficial qualities identified in *transformative* learning theory for peer-learning partnerships, namely "non-hierarchical status, non-evaluative feedback, voluntary participation, partner selection, authenticity, and establishment of mutual goals" (Taylor, 2008).

For this study, the degree to which the two community channels had to be utilised, for instance to overcome an impasse or clarify a point in the documentation, varied significantly between libraries. Some required multiple elaborate questions to be posted (e.g., *GeoViews, datashader*, *Altair*), and others (e.g., *Cartopy, GeoPandas, Plotly.py*) required no interaction. This could be due to a genuine gap in documentation , or insufficient user knowledge about the respective API. For this reason, this aspect was not considered as a viable comparative indicator.

The overall 'chances of success' in engaging with these informal channels also differed: Subject to the availability of community resources and time—note Taylor's (2008) reference to 'voluntary participation'—and assuming a user's question happens to match another user's prior knowledge, it is not uncommon that questions are resolved in a matter of hours. At other times, responses may never materialise or fail to solve a particular issue. During this study, both scenarios occurred, though the latter with much less frequency. It can be assumed that libraries with a larger user base (see Table S2 and S3) are more likely to provide timely support to new users.

## 5.2 Static visualisations

### 5.2.1 Documentation

**Table 3** summarises the documentation structure for each library in the static track as well as useful examples for implementing the visualisation task, producing categorical choropleth maps along with tiled base maps. API references are not listed.

| Library | Documentation structure | Useful examples + *respective source document* |
|---|---|---|
| GeoPandas | "Introduction" page, "User Guide", a nascent "Advanced Guide", and an "Examples" page | *Examples:* Categorical legend, adding tiled base maps |
| Cartopy | "Getting started" guide containing both API reference-style sections and examples, also a "Gallery" | *Getting started:* Matplotlib integration; *Gallery:* categorical legend, adding tiled base maps |
| geoplot | "Quickstart", "User Guide" and a "Gallery" | *User Guide:* Categorical legend and base map |
| Altair | "Getting started", "Tutorial", "User Guide", "Advanced Usage" and "Example Gallery" | *GitHub issue #588:* GDF to `mark geoshape`; *Tutorial:* choropleth maps, *User Guide:* using `alt.Scale` for imperative colormapping; |
| datashader | "Getting started" guide, "user guide", datashader-specific "Topics" and PyViz examples gallery covering datashader and other HoloViz products | *Getting started:* categorical transformation |
| GeoViews | "Introduction", "User Guide", "Gallery", "Topics" and PyViz examples gallery | *HoloViews User Guide:* categorical colormapping; *Gallery:* adding tiled base maps |

**Table 3:** Overview of library documentation structure and possible code examples

Newcomers may have to combine and adapt approaches from multiple sources, in some cases from related libraries, to implement a particular use case. The referenced *Altair* GitHub issue is an instance where examples useful for a particular use case are public though not (yet) included in the official documentation. All projects actively encourage users to submit 'pull requests', i.e. community-developed changes to the code base, to improve and expand on documentation (for example, Bednar 2018b).

*5.2.2 Code complexity*

Figure 2 compares the proxy measure for code complexity, the number of lines of code.

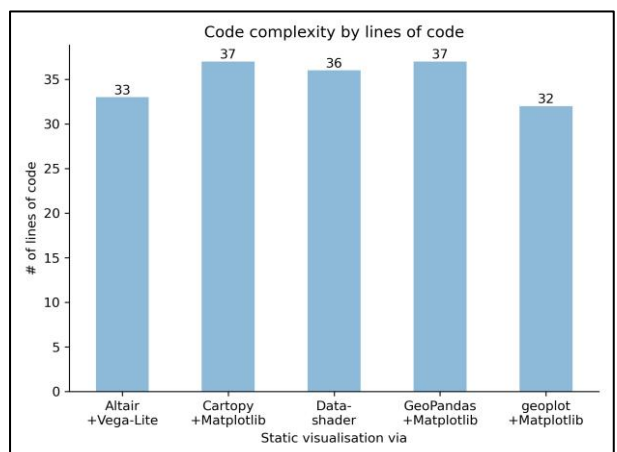

**Figure 2:** Static track code complexity

Code complexity is largely comparable. Generally, the use case in combination with available features has a significant impact on code verbosity. Especially with non-geospatial use cases from traditional data science, declarative approaches can be far more economical once key and value dimensions have been declared. Due to the imperative nature of the chosen visualisation task which required additional lines of code otherwise not necessary in a purely declarative approach, this relative advantage of declarative approaches is not apparent here.

*5.2.3 Adherence to map template and resource use*

Outputs using the complete dataset for all five implementations are presented in Figure 3. The panel descriptions indicate the main function or method through which the user interacts with the respective library.

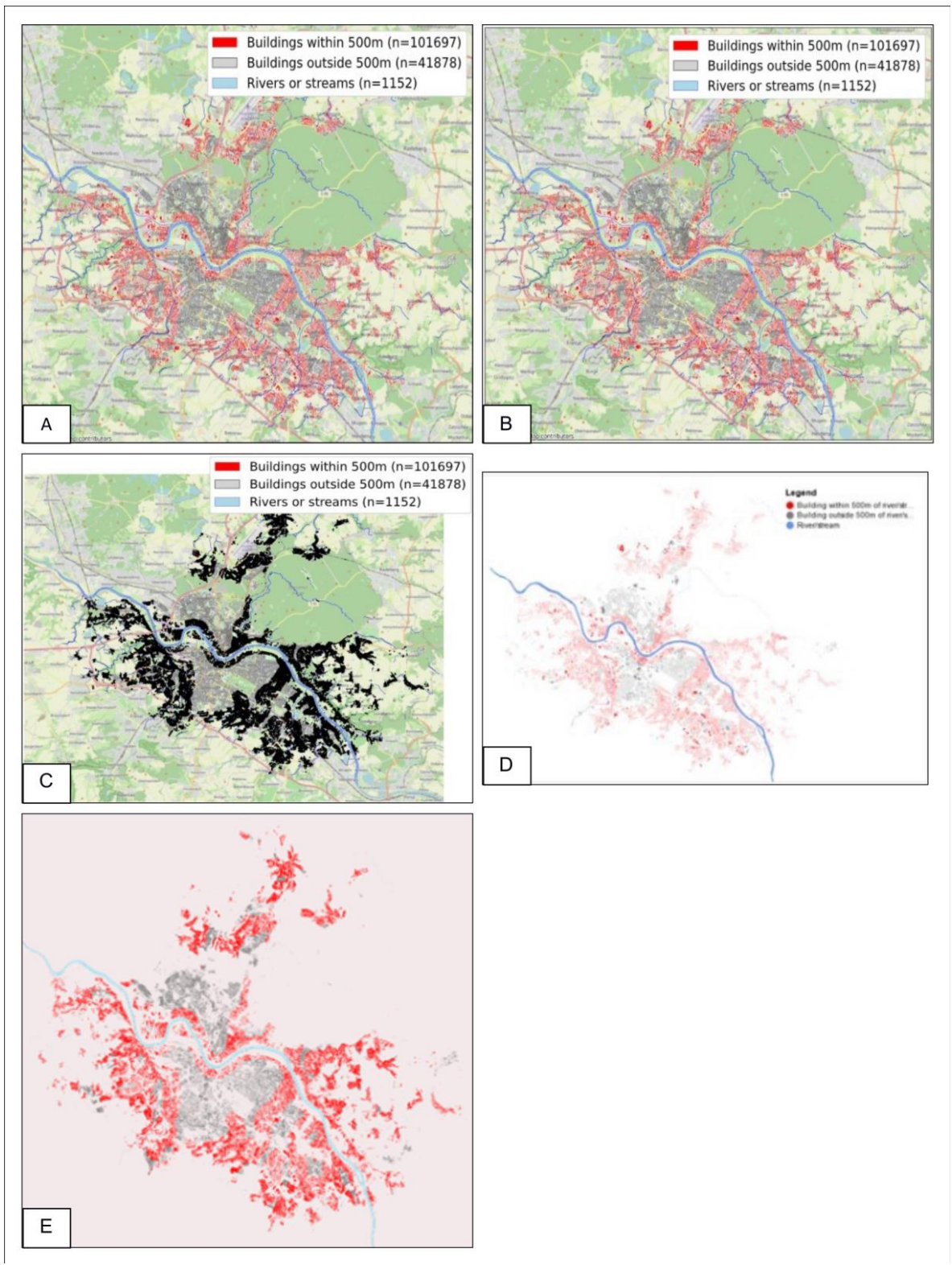

**Figure 3:** Static track comparison of outputs: (A) *GeoPandas*' `GeoDataFrame.plot()` method, (B) *Cartopy*'s `add_geometries()` function, (C) *Geoplot*'s `polyplot()` function, (D) *Altair*'s `Chart` class, (E) *Datashader*'s `transfer_functions.shade()` function

Key metadata and observations on the generated map products are summarised in **Table 4**.

| Library | Legend | Base map | Format | File Size[3] Subset | File Size[3] Complete | Comment on complete dataset output |
|---|---|---|---|---|---|---|
| *GeoPandas* + Matplotlib | Yes | Yes (*contextily*) | .svg | 2.2MB | 43.6MB | - |
| *Cartopy* + Matplotlib | Yes | Yes (*contextily*) | .svg | 0.9MB | 40.5MB | - |
| *geoplot* + Matplotlib | Yes | Yes (contextily) | .svg | 2.7MB | 68.7MB | Legend position slightly off-figure on the complete dataset, but rendering normally on the subset dataset. At low zoom levels, colours seem to diverge from *GeoPandas/cartopy* despite identical keyword arguments. |
| *Altair* + *Vega-Lite* | Yes | No | .png[4] | 61.8KB | 136KB | Base map support by Vega-Lite currently in development (see Vega-Lite GitHub issue #5758) |
| *datashader* | No | No | .png | 29.5KB | 103KB | *Included for illustration only:* No legends or axis elements when employed on its own. Main use case targets rasterisation/aggregation of large datasets prior to plotting, in tandem with core plotting libraries |

**Table 4:** Static track map elements and output metadata

### 5.2.4 CPU runtime

The mean CPU runtimes for the `renderFigure()` function measured using *cProfile* are presented in Figure 4.

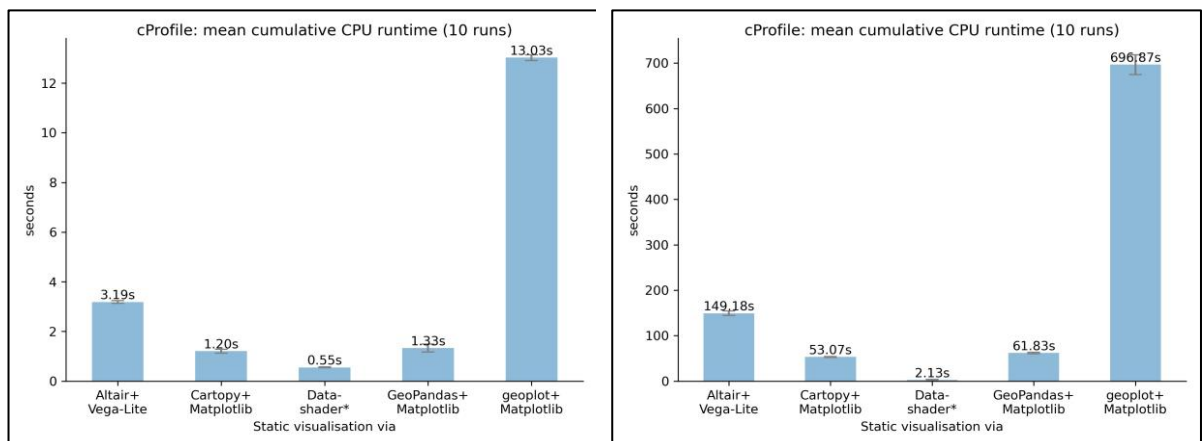

**Figure 4:** Static track performance benchmark for the subset (left) and complete dataset (right), error bars indicating the standard deviation

Of the three *Matplotlib*-based libraries, *geoplot*'s mean CPU runtime is consistent with the behaviour observed in the interpreter, but was only revealed through *cProfile* following the adjustment indicated in Table S3. Without the adjustment, *geoplot*'s average CPU runtime on the complete dataset stood at 36.29 seconds. The stark difference to, for instance, *Cartopy* may be due to different entry points to *Matplotlib*: *geoplot* initially calls the `matplotlib.pyplot.gca()` ('get current Axes') function, whereas *Cartopy*'s `add_geometries()` function instantiates *Matplotlib*'s `FeatureArtist` class. At the time of submission, despite contacting the respective developer communities and a review of the code base, it was not possible to determine why the performance of these *Matplotlib* entry points differed so significantly, suggesting an area for future research.

---

[3] This is the file size when saving the map product to disk.
[4] Saving to `.svg` is supported through the *Altair Saver* extension, though the package had reportedly not been tested for Windows and could not be successfully implemented prior to submission.

*Cartopy* marginally outperforms *GeoPandas,* while *Altair*'s interface to *Vega-Lite* ran three times longer than these two *Matplotlib* interfaces.

*Datashader*'s large standard deviation of 1.853 seconds with a mean of 1.564 seconds is due to its use of the *Numba* compiler, which provides similar performance to a traditional compiled language (Lam et al., 2015). The significant performance gains through *Numba* materialise only after compilation in the first run. This bears out in the individual *datashader* CPU runtimes observed, presented in Table S4.

## 5.3 Interactive visualisations

### 5.3.1 Documentation

| Library | Documentation structure | Useful examples |
|---|---|---|
| Bokeh | "First Steps", "User guide", "Gallery", interactive Binder Tutorial | User guide: mapping geo data and adding tiled base maps; First steps: adding a legend |
| GeoViews + Bokeh | *GeoViews*: "Introduction", "User Guide", "Gallery", "Topics" and PyViz examples gallery | HoloViews User Guide: overlays, using `group` for categorical colormapping; GeoViews User Guide: geometries; Gallery: adding tiled base maps |
| GeoViews + datashader + Bokeh | *HoloViews*: "Getting started" guide, "User guide", "Reference Gallery"; *GeoViews*: "Introduction", "User Guide", "Gallery", "Topics" and PyViz examples gallery; *datashader*: "Getting started" guide, "User guide", *datashader*-specific "Topics", PyViz examples gallery for all three libraries | HoloViews User Guide: working with large data; PyViz gallery: datashaded polygons + `inspect_polygons` enabling Bokeh hover tools; Bokeh User Guide: running a Bokeh Server |
| Plotly.py | "Getting started" and several "Examples" pages covering Plotly fundamentals and different chart types including maps | Gallery: map configuration and styling, adding tiled base maps to a choropleth map, discrete colormapping; StackOverflow: determining zoom levels for `choropleth_mapbox` |

**Table 5:** Overview of library documentation structure and possible code examples

The StackOverflow thread referenced in **Table 5** concerning *Plotly.py* is highlighted as illustrative of a general phenomenon when dealing with software, including open-source solutions: a non-obvious gap in documentation, requiring users to consult with the wider community through channels outside the official documentation. In this case, *Plotly.py's* documentation outlined a method for determining a suitable zoom level for its regular mapping functions such as `express.choropleth()` but does *not* mention that the same approach does not (yet) apply to another almost identical function, namely `express.mapbox_choropleth()`.

### 5.3.2 Code complexity

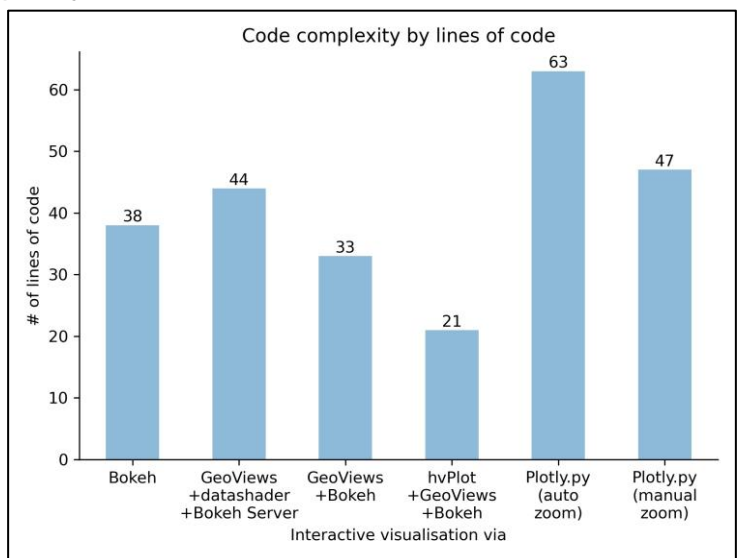

**Figure 5:** Interactive track code complexity

As shown in Figure 5, code complexity in the interactive track differed more widely.

When applied to this study's use case, *Plotly.py's* syntax was the most complex despite employing the more user friendly *Plotly Express* interface. This is due to the need to write an intermediary JSON file containing the geometries to disk (Plotly Technologies Inc., 2021). *Plotly.py*'s line count remains high even without adding auto zoom functionality.

*GeoViews* and *hvPlot* significantly simplify interacting with *Bokeh*'s standard API, with *hvPlot* offering the most succinct plotting interface of all short-listed libraries. These additional layers of abstraction may lead to a decrease in overall performance, however (see section 5.3.4).

*5.3.3   Adherence to map template and resource use*

Outputs using the complete dataset for all five libraries are presented in Figure 6, demonstrating their ability to present at least four data attributes simultaneously: two attributes representing geographic location, one attribute conveyed through symbology, and at least one additional attribute conveyed through the hover tool. The latter can be extended to display as many attributes as can comfortably fit the screen (as demonstrated in Figure 6 C, D).

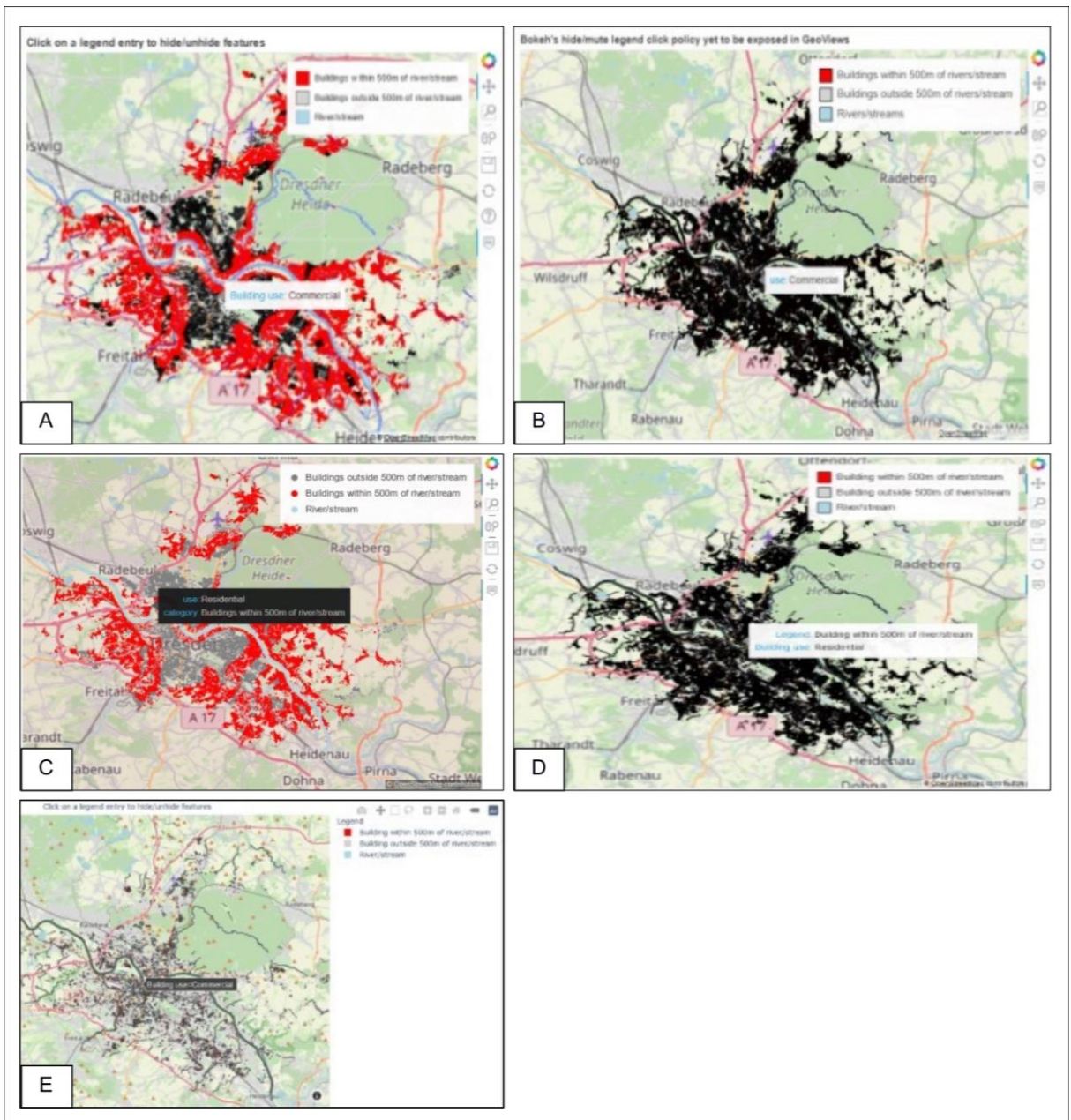

**Figure 6:** Interactive track output comparison, (A) *Bokeh*'s `figure.patches()` function, (B) *GeoViews*' `Polygons` class + *Bokeh* backend, (C) *GeoViews'* `Polygons` class + *datashader'*s `datashade` function + *Bokeh Server*, (D) *hvPlot*'s `hvplot()` method, (E) *Plotly.py*'s `express.choropleth_mapbox()` function

Key metadata and observations on the generated map products are summarised in **Table 6**.

| Library | Legend | Base map | Format | File Size | | Comment on complete dataset output |
| --- | --- | --- | --- | --- | --- | --- |
| | | | | Subset | Complete | |
| *Bokeh* | Yes | Yes | .html | 1.6MB | 70.4MB | Slow to respond on pan and zoom |
| *GeoViews* + *Bokeh* | Yes | Yes | .html | 2.1MB | 94.2MB | Slow to respond on pan and zoom. Colours differ from *Bokeh* at low zoom levels only. |
| *GeoViews* + *datashader* + *Bokeh Server* | Yes | Yes | .html | 864KB | 864KB | Rasterised polygons dynamically recalculated with every pan and zoom, highly responsive. Colours differ from *Bokeh* at low zoom levels only, possibly due to less overplotting. |
| *hvPlot* + *GeoViews* + *Bokeh* | Yes | Yes | .html | 1.1MB | 54.4MB | Slow to respond on pan and zoom. Colours differ from *Bokeh* at low zoom levels only. |
| *Plotly.py* | Yes | Yes | .html | 8.8MB | 247.2MB | Highly responsive on pan and zoom |

**Table 6:** Interactive track map elements and output metadata

The complete dataset presented challenges for almost all visualisation strategies, both during production and consumption of the map product.

There are large differences in file size, with *Plotly.py* being a clear outlier. Even though the output is highly responsive once fully loaded in the browser, the large file size makes it impractical for embedding into public-facing webpages. Online distribution would therefore require an on-demand download or prior map tile conversion.

While the file size for the *datashader*-based implementation is explained by *Bokeh* receiving, in essence, a raster image at a set resolution of the polygons currently visible in the viewport, the cause of the variation between the other three *Bokeh* implementations could not be ascertained. The *HoloViz* default to inline *BokehJS* instead of accessing it through a content delivery network, as is done by *Bokeh's* default API, only accounts for a marginal increase in file size. It also fails to explain why *GeoViews* and *hvPlot,* both *HoloViz* products, differ from *Bokeh*'s file size in opposite directions.

### 5.3.4 CPU runtime

The mean CPU runtimes for the `renderFigure()` function on the interactive track are presented in Figure 7.

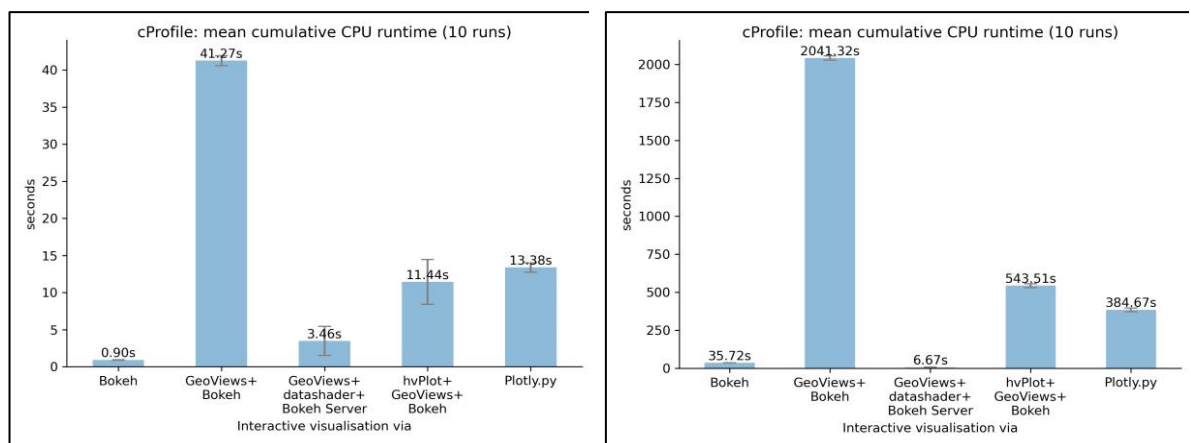

**Figure 7:** Interactive track performance benchmark for the subset (left) and complete dataset (right), error bars indicating the standard deviation

Stark differences in mean CPU runtime between the three *Bokeh*-based implementations making no use of *datashader* were observed: interfacing with *Bokeh* through *hvPlot* proved to be 3.7 times faster than interacting with *Bokeh* through *GeoViews*. Engaging with *Bokeh's* API directly outperformed *hvPlot* and *GeoViews* by a factor of 15.5 and 58, respectively. Despite consultations with the developers, it was not possible to ascertain the reasons for *GeoView's* apparently poor performance. One possible explanation is the additional layers of abstraction between *GeoViews* and *Bokeh* (via *HoloViews*), although *hvPlot* employs an even higher level of abstraction without incurring a similar penalty in performance.

The *GeoViews + datashader + Bokeh Server* implementation performed the best, combining *Bokeh's* interactive features with low client-side resource use. Runtimes in subsequent runs were reduced drastically (see Table S5), similar to what was demonstrated in Table S4.

# 6   Conclusions

## 6.1   Python for Smarter Cities: the give and take of open-source software

To put the study's results into a wider perspective, it must be emphasised that current library features, performance, comprehensiveness of documentation and support capabilities can, to a large extent, be attributed to funding availability, as well as time volunteered by contributors.

In 2015, the European Innovation Partnership on Smart Cities and Communities (EIP SCC) initiated a Memorandum of Understanding intent on avoiding proprietary and custom-built offerings, while promoting interoperability and common open standards (European Innovation Partnership on Smart Cities and Communities, 2015). This led to the DIN SPEC 91357:2017-12 standard, Reference Architecture for Model Open Urban Platforms (DIN SPEC 91357:2017-12). If cities are committed to a publicly-owned open data architecture, local authorities are not only in a position to enhance the sustainability of underlying software but should, in fact, be encouraged to do so. If the EIP SCC's vision is realised, open-source software applied to smart city initiatives is as much part of publicly-managed infrastructure as traditional hard assets, along with concomitant needs for investments in maintenance.

Organisations are already opting for supporting projects financially (NumFOCUS, Inc., 2021), either through unrestricted contributions or funding aimed at adding a particular functionality. Google, Microsoft and ESRI's support for the Geospatial Data Abstraction Library (GDAL, 2021c), the fundamental infrastructure underpinning the geospatial capabilities of the Python libraries presented here and many well-known commercial products, is one illustration of this. Alternatively, projects can be supported by allowing staff time to improve project documentation and functionality.

## 6.2   Conclusion

This study demonstrates that the Python ecosystem offers numerous options for generating engaging, consistently-formatted static and interactive map products. For some workflows, the value for local authorities compared to commercial desktop GIS applications is noteworthy: streamlined file management (Lin, 2012), high reproducibility (Ince et al., 2012; Lazer et al., 2014) that enhances institutional memory, and avoidance of licensing fees and vendor lock-ins.

The presented solutions' foundation in open source combined with their active development and a growing number of firms specialising in open-source GIS (e.g., QGIS.org, 2021) also means that unnecessarily bespoke, rather than multi-authority, shared solutions can be avoided (Copeland 2015) without having to forego professional support.

The long list review indicates that support for any geospatial data format is virtually guaranteed. In addition, all short-listed libraries were supported both by core teams of developers and an active user base.

The comparative measures presented offer initial assistance in selecting suitable candidates for specific user needs: for simple applications where ease of use is of greater importance than the details of symbology or the fine-tuning of map elements, high-level imperative or declarative approaches such as *hvPlot* or *Altair* provide clear advantages. On the other hand, interacting directly with *Matplotlib* or lower-level libraries such as *Bokeh* and *Plotly* offers a higher degree of customisability at the cost of a steeper learning curve.

Documentation and community support proved to be vital tools in flattening this curve. Formal documentation was of generally high quality across both tracks, providing a range of examples requiring only minor adjustments. Community support channels such as GitHub and StackOverflow were also invaluable resources.

The static track presented low variability in code complexity. This may be due to the imperative nature of the visualisation task, thus slightly disadvantaging declarative approaches. The interactive track had the largest variability in code complexity, ranging from the intuitive to the more verbose, requiring users to familiarise themselves with library-specific concepts. For the chosen use case, *hvPlot* offered the most succinct interface and *Plotly Express* the most complex, even when disregarding automatic zoom level detection as a necessary feature.

CPU runtime may be of minor importance when dealing with smaller datasets, with all short-listed libraries able to handle the subset dataset without difficulties. This was not the case for the full dataset. For static visualisations, *Cartopy's* interface to *Matplotlib*, closely followed by *GeoPandas'*, performed the best for both datasets, completing the rendering function almost three times faster than *Altair.* Several libraries showed significant differences despite all interfacing with *Matplotlib*, suggesting an area for further investigation.

For interactive visualisations, there were stark differences in mean CPU runtime between the three *Bokeh*-based implementations not relying on *datashader*. Engaging with *Bokeh's* API directly outperformed both interfacing via *hvPlot* and *GeoViews* by a factor of 15.5 and nearly 60, respectively. Combining *GeoViews, datashader* and *Bokeh Server* proved to be the most performant, highlighting the advantages gained through on-demand aggregation and transformation when handling large datasets.

It must be noted that with libraries relying on other complex technologies, some of which are not based on Python, it is possible that the study design did not establish a fully level playing field. Additional layers of abstraction between any core plotting library and the user-facing API are likely to lead to losses in performance, unless data aggregation and transformation strategies are applied as well.

In summary, Python-based visualisation offers local authorities powerful tools, free of vendor lock-in and licensing fees, to produce performant and consistently-formatted visualisations for both internal and public distribution.

**Glossary**

| *Declarative visualisation approach* | The user only specifies what should be rendered, with plotting details determined automatically. This separates specification from execution. |
|---|---|
| *Imperative visualisation approach* | The user manually specifies how something should be rendered, including individual plotting steps. Specification and execution are intertwined. |

**Funding**

This research did not receive any specific grant from funding agencies in the public, commercial, or not-for-profit sectors.

**Acknowledgements**

The authors would like to thank the developers and users of the libraries profiled here. Without their feedback and patience, this study would not have been possible.

**Declaration of interest: none**

# Supplementary material

## S1 Data and methods

*S1.1 System specifications and data source*

The system consisted of a Dell XPS 15 7590 with a 2.60GHz Intel® Core™ i7-9750H CPU and 16 GB of RAM running Windows 10. All scripts were run from within a Python v3.8.8. *conda* environment executed by the Visual Studio Code (VSCode) v1.59.1 Python Extension.

The dataset used in the analysis consisted of two tables (see Table S1) from a 14GB vector dataset containing the German city of Dresden's complete real-estate cadastre (Staatsbetrieb Geobasisinformation und Vermessung Sachsen, n.d.) covering an area of 637 square kilometres. The dataset is available for download as 111 individual files in the Extensible Mark-up Language (XML) format. The dataset's schema follows Germany's Authoritative Real Estate Cadastre Information System 'ALKIS' (Arbeitsgemeinschaft der Vermessungsverwaltungen der Länder der Bundesrepublik Deutschland, 2021), which itself is based on the ISO 19100 series of standards (ISO, 2021).

A production environment was emulated which, though Windows 10-based, otherwise relied entirely on open-source applications. Since municipalities are increasingly making use of PostgreSQL (Burger, 2019), and since Roy et al. (2019) further showed in-database queries to be substantially more efficient than in-memory queries through *GeoPandas*, two separate locally hosted PostGIS v3.0.0-enabled PostgreSQL v12.4 databases were populated, one containing the complete dataset and another containing a subset of geometries from only a single XML file. Table creation and the XML-to-PostgreSQL data transfer was performed using the *norGIS ALKIS Import* tool available via the advanced installation options of the *OSGeo4W* binary distribution for Windows (Open Source Geospatial Foundation, 2021).

Table 1 provides metadata for the two tables from the dataset which were utilised in the study.

**Table S7:** Spatial tables selected from the ALKIS dataset

| *Original table name* | Feature type | Geometry type | Feature count (complete) | Feature count (subset) |
|---|---|---|---|---|
| *AX_Gebaeude* | Buildings | CurvePolygon | 143,575 | 2,645 |
| *AX_Fliessgewaesser* | Rivers or streams | CurvePolygon | 1,152 | 12 |

*S1.2 Spatial query and common visualisation task*

As the spatial query itself was secondary to the research objective, the spatial query was kept simple and was performed via *GeoPandas'* `from_postgis()` function. Of the numerous columns available in each base table, only the feature ID and geometry columns were queried for both. However, to avail of interactive libraries' ability to indicate one or more additional data attributes through hover tools, a table update was performed in-database on the building tables' building use column translating the schema's four digit attribute values representing fine-grained building uses into broader, English language building use categories. This updated column was then also included in the query's SELECT statement.

Thus, the following three feature datasets were returned (subset feature counts in parenthesis):

a) 101,697 (1,729) building footprint polygons, including the building use column, completely within, or intersecting with, a 500m buffer around any river or stream;

b) 41,878 (916) building footprint polygons, including the building use column, completely outside said buffer;
c) 1,152 (12) river or stream polygons.

The resulting in-memory GeoDataFrames then directly or indirectly served as the input data for the selected libraries.[5] In the subsequent visualisations, a uniform symbology was applied to each of the returned feature sets in addition to adding, where possible, a categorical legend. For context and where libraries allowed for this functionality, the OpenStreetMap default layer was added as a base map, except during performance testing, thus producing a relatively consistently formatted 'map template' across both tracks and all libraries.

## S1.3 Library selection

To gain a first overview of candidate libraries, a long list of libraries informed by the list collated by PyViz.org (2019) was compiled including metadata such as:

1. Output formats (static images, interactive maps. or both);
2. general implementation strategy (e.g. a high-level interface to a third-party 'core' plotting library or a core library itself);
3. Installation channels and requirements (install via the *pip* package installer, the *conda* main channel or the *conda-forge* community channel, or via *setup.py*, optional extensions required);
4. Input formats and required data conversions;
5. Proxies suggesting the vibrancy of the developer and user community (*indicators*: number of GitHub releases since first release, number of total commits; date of last commit; number of contributors[6]; number of dependent packages and number of dependent repositories).

**Table S8:** Long list of libraries producing static visualisations

| Library[7] | Output formats | Interface / Architecture | Install | Geo input format/ conversion[8] | Community[9] |
|---|---|---|---|---|---|
| *GeoPandas* v0.9.0 (Jordahl et al., 2021) + *Matplotlib* - User Guide - API - GitHub | Identical to *Matplotlib*, i.e. wide range of vector and raster outputs for the non-interactive backend such as .svg and .png | High-level imperative interface with *Matplotlib* through *Pandas'* .plot() API. | conda conda-forge pip | Any Geographic Data Abstraction Library (GDAL)/OGR format, read via the *fiona* package | - 20 releases since 2014 - 1,396 commits - last commit in June 2021 - 141 contributors - 498 dep. packages - 8,149 dep. repositories ('repos') |
| *Cartopy* v0.18.0 (Met Office, 2021) + *Matplotlib* - User Guide - API - GitHub | Identical to *Matplotlib*, i.e. wide range of vector and raster outputs for the non-interactive backend such as .svg and .png | High-level imperative interface with *Matplotlib*. Extends *Matplotlib* Axes class as GeoAxes with projection capabilities specific to geographic applications | conda pip | Any GDAL/OGR format via *GeoPandas* | - 33 releases since 2012 - 2,281 commits - last commit in May 2021 - 86 contributors - 152 dep. packages |

---

[5] The processing time for the data acquisition function which included PostGIS processing and generating the GeoDataFrames in-memory, ranged from 0.5 seconds for the subset to 1 minute and 55 seconds for the complete dataset.
[6] It should be noted that most contributors listed on GitHub are irregular one-time contributors, whereas the team of full-time 'core' developers is substantially smaller. For instance, *Bokeh* currently lists two 'core team' members (Bokeh 2021) but has 504 contributors on Github. This measure should therefore only be seen as indicative of the project's community appeal.
[7] Only libraries on the short-list of either track are referenced and listed with specific version numbers.
[8] Several libraries can consume a wide range of non-geospatial data containers as well. For brevity, only umbrella terms for primarily geospatial input formats are listed here and in **Table S9**.
[9] GitHub data for the static track are as of June 2021.

| Library | Output formats | Interface / Architecture | Install | Geo input format/ conversion | Community |
|---|---|---|---|---|---|
| | | | | | - 1,523 dep. repos |
| geoplot v0.4.3 (Bilogur et al., 2021) + Matplotlib<br>- User Guide<br>- API<br>- GitHub | Identical to Matplotlib, i.e. wide range of vector and raster outputs for the non-interactive backend such as .svg and .png | Very high-level imperative interface with Matplotlib | conda | Any GDAL/OGR format via GeoPandas, data in EPSG:4326 | - 17 releases since 2017<br>- 373 commits<br>- last commit in May 2020<br>- 8 contributors<br>- 2 dep. packages<br>- 245 dep. repos |
| Altair v4.1.0 (VanderPlas et al., 2018) + Vega-Lite (Satyanarayan et al., 2017)<br>- User Guide<br>- API<br>- GitHub | .json, .vl.json, .vg.json, .html, .png, .svg, .pdf | High-level declarative interface based on Vega-Lite. | conda-forge<br>pip<br>ext.: Altair Viewer and Altair Saver | Any GDAL/OGR format via GeoPandas, data in EPSG:4326, then conversion to_json and deserialisation to Python object via json.loads | - 23 releases since 2016<br>- 3,065 commits<br>- last commit in April 2021<br>- 116 contributors<br>- 166 dep. packages<br>- 11,914 dep. repos |
| datashader v0.13.0 (Bednar et al., 2021)<br>- User Guide<br>- API<br>- GitHub | .png, though without axes, legends, or other plot annotations | Aggregates and rasterizes datasets of any size, avoiding common plotting pitfalls (overplotting, oversaturation, undersampling, etc.). Meant to be used in conjunction with GeoViews/HoloViews + Matplotlib/Bokeh/Plotly to add map elements (legends, axes, base maps etc.) | conda<br>pip | Any GDAL/OGR format via GeoPandas, then conversion to SpatialPandas GeoDataFrames | - 84 releases since 2016<br>- 1,295 commits<br>- last commit in June 2021<br>- 41 contributors<br>- 47 dep. packages<br>- 661 dep. repos |
| GeoViews v1.9.1 (Rudiger et al., 2021a) + Matplotlib<br>- User Guide<br>- GitHub<br>- v1.9.1 | Identical to Matplotlib, i.e. wide range of vector and raster outputs for the non-interactive backend such as .svg and .png | High level HoloViews-based (Rudiger et al., 2021b) declarative interface to Matplotlib, incorporating Cartopy's projection support. Categorical legends apparently not yet supported when used in conjunction with Matplotlib. | conda | Any GDAL/OGR format via GeoPandas. | - 52 releases since 2016<br>- 656 commits<br>- last commit in April 2021<br>- 22 contributors<br>- no dependents listed |

**Table S9:** Long list of libraries producing interactive visualisations

| Library | Output formats | Interface / Architecture | Install | Geo input format/ conversion | Community[10] |
|---|---|---|---|---|---|
| Bokeh v2.3.2 (Bokeh Development Team, 2018)<br>- User Guide<br>- API<br>- GitHub | .html (static: .png, .svg) | Choice of high-level interfaces for data scientists or low-level interfaces for developers requiring more control. Builds on its own Python and JS library to render the data in HTML/JS. Optional WebGL support to make use of hardware acceleration using graphics processing units (GPUs). Currently no WebGL support for Patches glyphs as used in this study. | conda<br>pip | Any GDAL/OGR format via GeoPandas, then to GeoJSON, then to Bokeh's GeoJSON-DataSource class | - 91 releases since 2013<br>- 19,090 commits<br>- last commit in June 2021<br>- 504 contributors<br>- 582 dep. packages<br>- 34,140 dep. repos |
| Plotly.py v4.14.3 (Plotly Technologies Inc. 2015)<br>- User Guide<br>- API<br>- GitHub | .html, (static: .png, .jpg, .jpeg, .svg, .pdf, .webp, .eps) | High-level declarative charting library on top of plotly.js. Well-integrated with dash. Includes an even higher level API, Plotly Express, utilised in this study. | conda<br>pip | Any GDAL/OGR format via GeoPandas plus on-disk GeoJSON | - 113 releases since 2013<br>- 5,046 commits<br>- last commit in June 2021<br>- 162 contributors<br>- 1 dep. package<br>- 5 dep. repos |
| GeoViews v1.9.1 + Bokeh v2.3.2<br>- User Guide<br>- GitHub | HTML | HoloViews-based high-level interface to Bokeh (or, alternatively, to Plotly), without applying datashader, thus more suitable for smaller datasets | conda | Any GDAL/OGR format via GeoPandas | [multi-library] |
| GeoViews v1.9.1 + | HTML | Single high-level declarative interface to Bokeh backend and Bokeh Server, | conda<br>pip | Any GDAL/OGR format via | [multi-library] |

---

[10] GitHub data for interactive track as of June 2021, except for hvPlot (August 2021).

| | | | | | |
|---|---|---|---|---|---|
| *datashader* v0.13.0 + *Bokeh* v2.3.2<br>- User Guide 1 and 2<br>- API<br>- GitHub | | with *datashader* dynamically serving rasterised polygons, suitable for large datasets. Incorporates *Cartopy*'s projection support. | | *GeoPandas* followed by conversion to *SpatialPandas* GeoDataFrames | |
| *hvPlot* v0.7.3 (Rudiger et al., 2021c) + *Bokeh*<br>- User Guide<br>- GitHub | HTML | A high-level plotting API for the PyData ecosystem built on top of *HoloViews*, with interactivity provided through *Bokeh*. Incorporates *Cartopy*'s projection support via *GeoViews*. | conda<br>pip | Any GDAL/OGR format via *GeoPandas* | 87 releases since 2018<br>- 419 commits<br>- last commit in August 2021<br>- 30 contributors<br>- 38 dep. packages<br>- 667 dep. repos |
| *Altair* v4.1.0 + *Vega-Lite*<br>- User Guide<br>- API<br>- GitHub | N/A | High-level declarative interface to *Vega-Lite.* As yet no *Vega-Lite* support for interactive geographic data (see Altair GitHub issue #679). | conda-forge<br>pip | N/A | 23 releases since 2016<br>- 3,065 commits<br>- last commit in April 2021<br>- 116 contributors<br>- 166 dep. packages<br>- 11,914 dep. repos |
| *folium*<br>- User Guide<br>- API<br>- GitHub | HTML | Allows for adding a range of map elements to *Leaflet.js* base maps including interactive markers, geometry layers, and legends. | conda-forge<br>pip | Any GDAL/OGR format via *GeoPandas,* then conversion to GeoJSON | 21 releases since 2014<br>- 1,474 commits<br>- last commit in March 2021<br>- 118 contributors<br>- 145 dep. packages<br>- 10,257 dep. repos |
| *mplleaflet*<br>- GitHub | HTML | Very high-level interface converting static *mpl*-produced plots into interactive Leaflet maps. | pip | Any GDAL/OGR format via *GeoPandas.* | 4 releases since 2015<br>- 84 commits<br>- last commit in March 2018<br>- 10 contributors<br>- 15 dep. packages<br>- 308 dep. repos |
| *geoplotlib*<br>- API<br>- GitHub | Rasterised vector data displayed in an interactive window or IPython notebook on top of OSM tile map. | Builds onto *numpy/ scipy* for numerical computations and OpenGL/*pyglet* for rendering. Reportedly suitable for large datasets (Cuttone et al., 2016) | pip<br>setup.py | .shp, .csv, pandas dataframe with geographic coordinates directly through *geoplotlib*, or prior conversion of GeoDataFrames to JSON | no releases published<br>- 159 commits<br>- last commit in May 2019<br>- 8 contributors<br>- 1 dep. package<br>- 101 dep. repos |

*S1.4 Measures taken to improve comparability of cProfile measurements*

The following measures were taken to improve comparability of results:

1. The Python kernel was restarted before each new benchmarking session;
2. To prevent some libraries, such as *Cartopy*, from re-using an already drawn canvas, each run was executed individually rather than automatically as part of a `for` loop (with the exception of the runs of *datashader,* see **Table S11** and **Table S12**);
3. During performance measurement, no base map tiles were added and figures were not written to disk;
4. To account for some libraries' lazy execution of underlying rendering functions, to force rendering within the interactive interpreter window during the course of the function call, or prevent libraries from displaying the figure in a browser window by default, the adjustments outlined in **Table S10** were added to respective scripts depending on libraries' particular behaviour if central calls to, say, a `plot` or `chart` object are made within a Jupyter Notebook/the VSCode Python Extension and as part of a function call:

**Table S10:** Usage adjustments to enforce rendering within VSCode as part of function call

| Static track | | Interactive track | |
|---|---|---|---|
| **Library** | **Adjustment** | **Library** | **Adjustment** |
| GeoPandas + Matplotlib | `fig.canvas.draw()` (pro-forma only, no effect on behaviour or performance) | Bokeh | `bokeh.io.output.output_notebook()` … `bokeh.io.show(plot)` |
| Cartopy + Matplotlib | `fig.canvas.draw()` | GeoViews + Bokeh | `bokeh.plotting.show(gv.render(plot))` |
| geoplot + Matplotlib | `matplotlib.pyplot.gcf()` | GeoViews + datashader + Bokeh Server | None |
| Altair + Vega Lite | `altair.renderers.enable('mimetype')` … `IPython.display.display(chart)` | hvPlot + Bokeh | `IPython.display.display(plot)` |
| datashader | None | Plotly.py | None |

## S2 Results

*S2.1 Individual CPU runtimes involving datashader*

**Table S11:** Individual CPU runtimes in seconds for *datashader* (subset dataset)

| run # | time | run # | time |
|---|---|---|---|
| 1 | 6.837 | 6 | 0.993 |
| 2 | 0.946 | 7 | 1.000 |
| 3 | 0.979 | 8 | 0.967 |
| 4 | 1.032 | 9 | 0.961 |
| 5 | 0.972 | 10 | 0.951 |

It should be noted that the runtimes in **Table S12** only refer to the initial plot creation. While running a local Bokeh Server, rasterised polygons were generally perceived to update on zoom and pan in the sub-second range.

**Table S12:** Individual CPU runtimes in seconds for plot creation using *GeoViews + datashader + Bokeh Server* (complete dataset)

| run # | time | run # | time |
|---|---|---|---|
| 1 | 11.354 | 6 | 6.147 |
| 2 | 6.380 | 7 | 6.256 |
| 3 | 5.877 | 8 | 6.028 |
| 4 | 5.720 | 9 | 6.566 |
| 5 | 5.982 | 10 | 6.392 |

## Supplementary References